# Incompatibility between the principle of the constancy of the speed of light and the Lorentz contraction in the GPS Experiment

Masanori Sato

*Honda Electronics Co., Ltd.,*
*20 Oyamazuka, Oiwa-cho, Toyohashi, Aichi 441-3193, Japan*

**Abstract:** Incompatibility between the principle of the constancy of the speed of light and the Lorentz contraction in the global positioning system (GPS) is discussed. The GPS works precisely in the earth-centered locally inertial (ECI) coordinate system on the condition that the speed of light $c$ is assumed to be constant regardless of the inertial motion of the GPS satellites and the earth. The inertial system of the earth travels not only in the solar system at the velocity 30 km/s but also in the cosmic microwave background (CMB) at the velocity 700 km/s. The deviation on the car navigation system by the Lorentz contraction of 700 km/s is maximally estimated as 54 m. However, such a large deviation is not observed: that is, the Lorentz contraction is not observed in the gravitational field of the earth. If there is a Lorentz contraction, the GPS cannot work so precisely. The GPS satellites are in the gravitational field of the earth, therefore the system should be interpreted by the theory of general relativity as well as special relativity.

1. Introduction

Ashby [1] described that the principle of the constancy of the speed of light, $c$, finds application as the fundamental concept on which the GPS is based. The global positioning system (GPS) works in the earth-centered locally inertial (ECI) coordinate system. The atomic clocks in the GPS satellites show the effect of the theory of special and general relativity. A car navigation system works very precisely. As far as the assumption that "the speed of light, $c$, is constant" is concerned, there is no effect neither of the velocity, $v_E$=700 km/s, in the CMB, nor of the 30 km/s in the solar system. The assumption that the speed of light, $c$, is constant is completely proved in the GPS experiments [1].

At the same time, the GPS experiments also show that there is no Lorentz contraction. That is, the assumption that the speed of light $c$ constant is incompatible with the Lorentz contraction. If the Lorentz contraction is calculated using the velocity $v_E$=700 km/s, the deviation of the car navigation system is maximally estimated as 54 m. There is no such large deviation experimentally detected. Thus we can conclude that we should interpret that there is no Lorentz contraction in the GPS.

As mentioned in the previous report [2], I did not know the reason why the earth-centered earth-fixed (ECEF) frame or the ECI coordinate system works well. I considered that it may be caused by the Lense-Thirring effect, which is an effect that is produced when the gravitational field of the earth pulls an inertial system. Thirring predicted that a rotating massive body would pull the surrounding space-time. This effect was predicted in the early stages of general relativity, and was checked in an experiment done in recent years [3]. It seems that the gravitational field of the earth pulls the surrounding inertial system by an effect similar to the Lense-Thirring effect. This similar effect is not a rotation but rather a linear dragging of the inertial system. However, there is no discussion of an inertial frame dragging. **Figure 1** illustrates the concept of an earth-centered local gravitational system which moves with the permittivity ε and permeability μ. At this stage, I consider that either the ECEF frame or the ECI coordinate system works correctly under the consideration of the theory of general relativity which contains the Lense-Thirring effect.

In this report, interpretation of the Lorentz contraction in the GPS experiment is discussed: that is, in a moving gravitational system, there is no Lorentz contraction detected.

2. Interpretation of the speed of light in the GPS experiment

The inertial system is constructed by the GPS satellites and the earth. **Figure 1** shows that the GPS satellites and the earth drift in the CMB at the velocity 700 km/s. The distance between the GPS satellites and a car on earth is measured using the signals from the GPS satellites. **Figure 2** shows a coded signal which contains the information of the satellite's position on the orbit and the precise time of the atomic clock in the satellite. If the signals are detected by a car on earth, the distance between the satellite and the car is calculated as $300,000(km/s) \times t_D(s)$. Using the four satellites the position of the car on earth is identified [1]: the position is calculated precisely using the constant value of the speed of light *c*. Therefore the speed of light *c* is concluded to be constant regardless of the motion of the inertial system.

In the GPS, atomic clocks in the GPS satellites can be treated by the ECI coordinate frame: the atomic clocks work accurately in the ECI coordinate frame, using the theories of special and general relativity [1].

**Figure 1** also shows a new assumption that the local gravitational field drags the surrounding space-time (light blue part) of the earth. It is similar to the acoustic waves in the atmosphere as will be discussed in section 3. **Figure 1** illustrates that the isotropic constancy of the speed of light is caused by the theory of general relativity: that is the gravitational field drags the space-time. (At this stage, I have not carried out any theoretical considerations.)

The GPS satellites travel in the gravitational field at the velocity of 4 km/s. The GPS satellites are traveling as if they are in the absolute stationary frame. The effects of the theory of special relativity cause a 7.1 μs time delay every day. This indicates that the GPS satellites do not drag the permittivity ε and permeability μ. There is the gravitational effect which causes a 45.7 μs time gain every day compared with an atomic clock on earth.

3. Analogy of electromagnetic and acoustic waves

In Maxwell's equation, the phase velocity of an electromagnetic wave, *c*, is described as

$$c = \frac{1}{\sqrt{\varepsilon \mu}}. \tag{1}$$

Where $\varepsilon=\varepsilon_r\varepsilon_0$, $\mu=\mu_r\mu_0$, $\varepsilon_0$ is the permittivity and $\mu_0$ is the permeability of free space, and $\varepsilon_r$ is the relative permittivity and $\mu_r$ is the relative permeability. If the permittivity, $\varepsilon$, and permeability, $\mu$, in the local gravitational field of the earth move with the earth, the speed of light, *c*, is constant regardless of the motion of the inertial system of the GPS satellites and the earth. It is very similar to the acoustic waves in the atmosphere. The phase velocity of acoustic waves is described as

$$c_A = \sqrt{\frac{C}{\rho}}, \tag{2}$$

where C is the coefficient of stiffness and $\rho$ is the density. C and $\rho$ move with the earth because C and $\rho$ are characteristic of the atmosphere which moves with the earth. The sound source, for example a car horn, does not move with the atmosphere.

There is an analogy between electromagnetic and acoustic waves in that the phase velocity is defined by the coefficients. Acoustic waves in the atmosphere travel at the sound velocity of 340 m/s independent of the motion of earth in the solar system. Electromagnetic waves in the gravitational field of the earth are similar to acoustic waves in the atmosphere.

Table 1 Comparison of electromagnetic and acoustic waves

|   | Waves | Phase velocity | Coefficients |
|---|---|---|---|
| 1 | Electromagnetic wave | $c = \dfrac{1}{\sqrt{\varepsilon \mu}}$ | $\varepsilon$: permittivity<br>$\mu$: permeability |
| 2 | Acoustic wave | $c_A = \sqrt{\dfrac{C}{\rho}}$ | C: coefficient of stiffness<br>$\rho$: density |

GPS satellites are similar to a car horn on the moving car: that is, the air around the car does not move with the car. While this is a very simple and rough illustration, it is, however, intuitive.

4. Deviation on the car navigation system by the Lorentz contraction

If we assume that the absolute stationary state is the CMB, the Lorentz contraction for $v_E$=700 km/s is calculated as follows,

$$\frac{1}{\sqrt{1-\left(\frac{v_E}{c}\right)^2}} = \frac{1}{\sqrt{1-\left(\frac{700}{300,000}\right)^2}} = 0.9999973. \tag{3}$$

Thus the Lorentz contraction is $1 - 0.9999973 = 2.7 \times 10^{-6}$.

**Figure 3** shows a critical condition of the GPS satellites and a car on earth: the direction of the GPS satellite seen from the car is parallel to $v_E$=700 km/s. (In this case, the CMB is assumed to be the absolute stationary state.) The distance between the GPS satellite and the car is around 20,000 km. Thus the deviation on the car navigation system by the Lorentz contraction is maximally estimated as,

$$20,000 km \times (1 - \frac{1}{\sqrt{1-\left(\frac{v_E}{c}\right)^2}}) = 20,000 km \times 2.7 \times 10^{-6} = 0.054 km = 54 m. \tag{4}$$

Table 2 Deviation on the car navigation system by the Lorentz contraction

| Background | Velocity of the inertial system | Deviation on the car navigation system by the Lorentz contraction | Periodic dependence |
|---|---|---|---|
| CMB | 700 km/s | 54 m | Sidereal time |
| Solar system | 30 km/s | 0.1 m=100 mm | Solar time |

In the solar system, where $v_E$=30 km/s in equation (4), the deviation is only 100 mm. Table 2 shows the deviation on the car navigation system calculated from the Lorentz contraction. In the interferometer of L1 band (1575.42 MHz, wavelength: 190 mm), relative accuracies of millimeters are reported [1]. This is the precise positioning analysis with carrier-phase measurements.

If, assuming that the CMB is the absolute reference frame, there is no deviation on the car navigation system, we can conclude that there is no Lorentz contraction. If the gravitational field of the earth pulls an inertial system, the ECI coordinate system works. As discussed previously, it is similar to sound propagation in the atmosphere on earth.

5. Discussion
5.1 The constancy of the speed of light

This report starts from the simple question of why the speed of light, *c*, has isotropic constancy in the GPS. The earth moves in the universe, therefore the Lorentz contraction is expected. However, the GPS experiment shows that there is no evidence of the Lorentz contraction. The earth seems to be in the absolute stationary state.

To solve this problem, using the analogy of acoustic waves, gravitational fields of the earth moves with the permittivity, ε, and permeability, μ. The relative permittivity, $ε_r$, seems to move with the earth, however, I do not have any idea of the motion of the permittivity, $ε_0$, and permeability, $μ_0$. According to the GPS experimental data, the illustration in **Fig. 1** is acceptable. This assumption of motion of the permittivity, ε, and permeability, μ, with the earth makes the reference time of the earth be coincident with that of the absolute stationary state after excluding the gravitational effect of the earth. A clock in the gravitational field of the earth is not affected by the drift motion in the CMB.

The Michelson-Morley experimental results are easily explained. This is because the gravitational local fields are equivalent to the absolute stationary state: that is, the two light paths of the Michelson-Morley experiment are completely equivalent. Thus the motion of the earth cannot be detected. The illustration in **Fig 1** is intuitive.

5.2 The frame dragging effect

The ECI coordinate system appears to work only in the gravitational field but not in free space. In this report, the gravitational field is assumed to drag the space-time around the earth. The dragging of space indicates that the permittivity, ε, and permeability, μ, move with the gravitational field. The dragging of the relative permittivity, $ε_r$, and permeability, $μ_r$, can be interpreted using the analogy of the acoustic waves in the atmosphere. This is because the relative permittivity, $ε_r$, and permeability, $μ_r$, adhere to the atmosphere.

At this stage, the frame dragging is discussed in rotating or accelerated motions: there is no discussion of an inertial frame dragging. Although I have no idea of the dragging of $ε_0$ and $μ_0$, however, the inertial frame dragging with the gravitational field is reasonable.

The dragging of time indicates that the reference time is not affected by the velocity $v_E$: that is the reference time does not need to consider the drift motion of the inertial frame. Only the gravitational field affects the reference time. That is time dilation by gravity.

6. Conclusion

The GPS, which constructs the local gravitational system, is considered in the absolute stationary state. The isotropic constancy of the speed of light, *c*, is sustained. There is no Lorentz contraction detected.

The GPS precisely works in the ECI coordinate system. If only the theory of special relativity is applied to the interpretation of the GPS experiment, the deviation on the car navigation system by the Lorentz contraction of 700 km/s in the CMB is maximally estimated as 54 m. However, there is no such deviation. The gravitational field drags the relative permittivity, $ε_r$, and permeability, $μ_r$, around the earth. Furthermore, the theory of general relativity should be taken into consideration. The GPS should be interpreted by the theory of general relativity as well as special relativity.

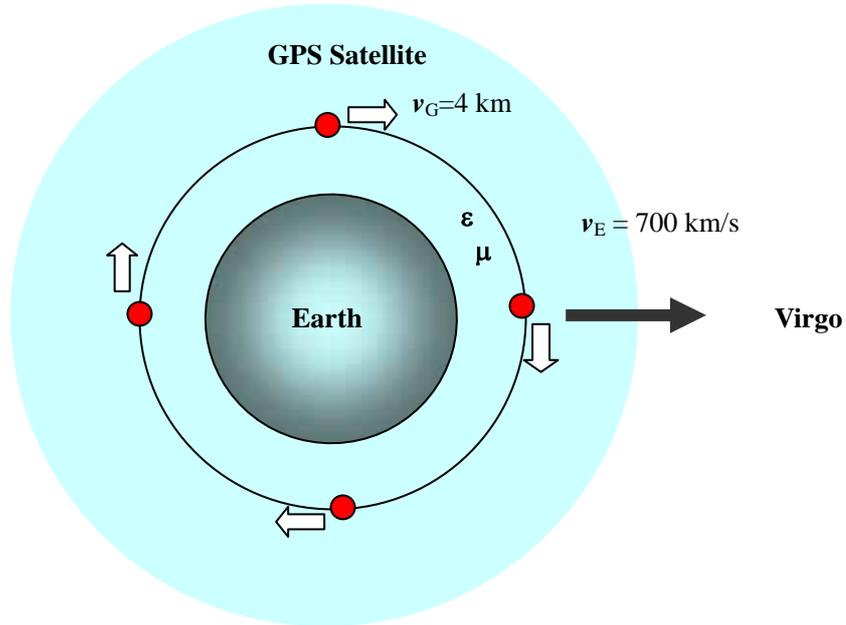

**Fig. 1**  Motion of the earth and the GPS satellites in the CMB
The earth and the GPS satellites move towards the constellation Virgo in the CMB at the velocity of 700 km/s. It is assumed that the permittivity ε and the permeability μ around the earth move with the gravitational field of the earth. The gravitational field, which is drawn as light blue in **Fig. 1**, drags the space-time around the earth. The GPS satellites move around the earth at the velocity $v_G$=4 km/s around 20,000 km from the ground level. The GPS satellites do not drag the space-time.

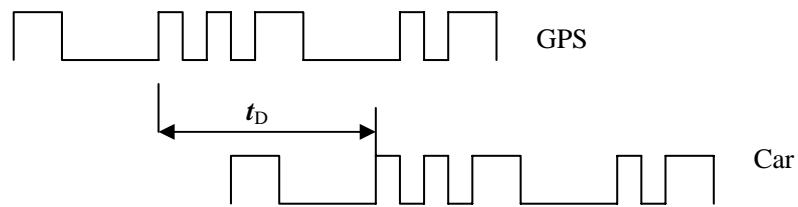

**Fig. 2**  Coded signal of the GPS satellite and the detected signal by a car navigation system on earth. The distance between the GPS satellite and the car is calculated using the time delay $t_D$ as
$300,000(km/s) \times t_D(s)$.

In the GPS, the time delay $t_D$ is measured precisely because the distance critically depends on $t_D$. In the GPS experiment, the assumption that the speed of light, $c$, is constant regardless of the velocity of the inertial system is confirmed.

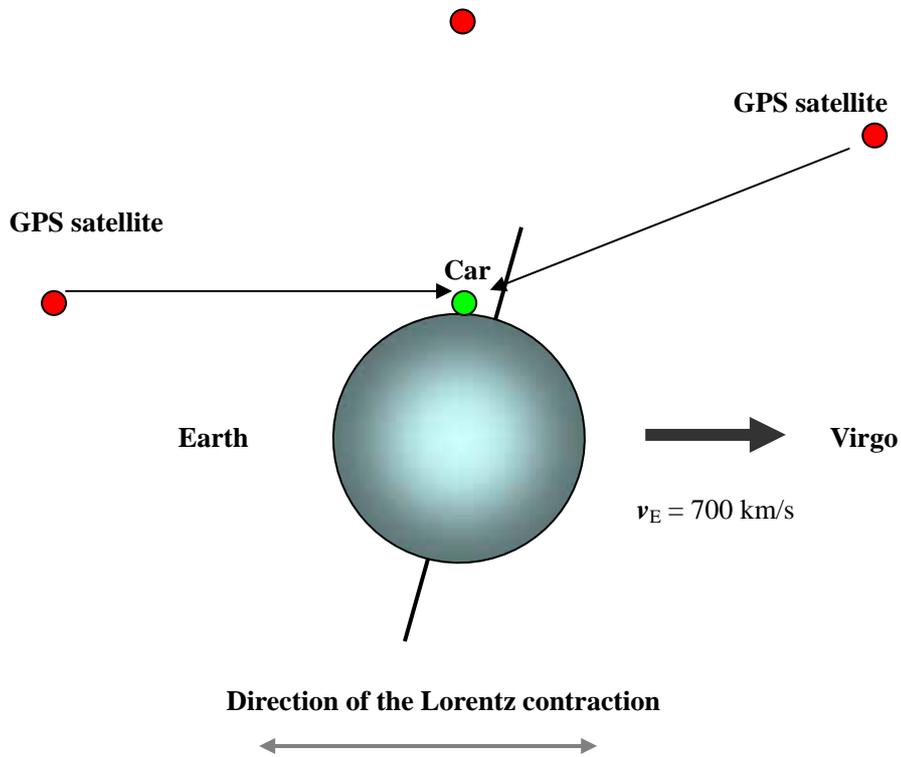

**Fig. 3** Illustration of the earth motion in the CMB. The earth motion is assumed to be at some 700 km/s towards the constellation Virgo. At a moment when a car on earth faces two GPS satellites, the car detects the speed of light *c* from two GPS satellites. According to the theory of special relativity, the Lorentz contraction of $2.7 \times 10^{-6}$ occurs in the direction of $v_E$. However, there is no Lorentz contraction detected in the GPS.